\documentclass[12pt,twoside]{amsart}
\usepackage{amssymb}
\usepackage[english]{babel}
\usepackage[utf8]{inputenc}
\usepackage{algorithm}
\usepackage[noend]{algpseudocode}
\usepackage{natbib}
\usepackage{comment} 
\usepackage{tikz}
\usetikzlibrary{calc}
\usepackage{verbatim}
\usepackage{subcaption}
\usepackage[labelformat=parens,labelsep=quad,skip=3pt]{caption}
\usepackage{graphicx}
\usepackage{enumerate}

\usepackage{kantlipsum} 
\def\RR{\mathbb{R}}    
\def\PP{\mathbb{P}}    
\def\EE{\mathbb{E}}    

\def\proof{\noindent{\em Proof.}~}
\def\eproof{\mbox{\ }\hfill$\square$}
\newtheorem{theorem}{Theorem}
\newtheorem{lemma}{Lemma}
\newtheorem{proposition}{Proposition}
\newtheorem{corollary}{Corollary}

\newtheorem{definition}{Definition}

\newtheorem{example}[theorem]{\textsc{Example}}
\newtheorem*{example*}{Example}

\date{}

\title{Censored Beliefs and Wishful Thinking}
\author{Jarrod Burgh\quad \quad Emerson Melo}

\date{\today}

\calclayout
\calclayout
\thanks{Department of Economics,
	Indiana University, Bloomington, IN 47408, USA. Email: {\tt \{jburgh,emelo\}@iu.edu}. We are very grateful to Duarte Goncalves, Marco Acosta, Jack Berger, Collin Raymond, Ben Bushong,  and participants at SAET 2023 and Georgia-Tech MWET 2023, for their valuable comments and suggestions that have greatly improved the paper. This paper, previously circulated under the title ``\emph{Wishful Thinking is Risky Thinking: A Statistical-Distance Based Approach},'' has been divided into two separate and complementary works: ``\emph{Wishful Thinking is Risky Thinking}'' and ``\emph{Censored Beliefs and Wishful Thinking.}'' Emerson Melo thanks the funding provided by the Institute for  Advanced Studies (IAS) at Indiana University-Bloomington through the  ``\emph{Recently Tenured Working Group}" internal grant.  } 
 
\begin{document}
		\begin{abstract}  
We present a model elucidating wishful thinking, which comprehensively incorporates both the costs and benefits associated with biased beliefs. Our findings reveal that wishful thinking behavior can be characterized as equivalent to superquantile-utility maximization within the domain of threshold beliefs distortion cost functions. By leveraging this equivalence, we establish WT as driving decision-makers to exhibit a preference for choices characterized by skewness and increased risk. Furthermore, we discuss how our framework facilitates the study of optimistic stochastic choice and optimistic risk aversion.
\end{abstract}

	\maketitle
	
	\pagestyle{myheadings} \thispagestyle{plain} \markboth{ }{ }

	\vspace{3ex}
	\small
\noindent {JEL classification: D01, D80, D84} \\
	{\bf Keywords:} wishful thinking, risk, optimism, quantile maximization, preference for skewness

	\thispagestyle{empty}
	
	\newcommand{\spacing}[1]{\renewcommand{\baselinestretch}{#1}\large\normalsize}
	\textwidth      5.95in \textheight 600pt
	\spacing{1.1}
	\newpage
	\section{Introduction}\label{Intro}
Wishful thinking (WT) behavior refers to the inclination to overestimate the probability of favorable events while underestimating the likelihood of unfavorable events (\cite{Aue_et_al_2012}). 

\smallskip 

There is substantial evidence of WT behavior in economic decision-making, with studies spanning various domains to illustrate this phenomenon. For instance, \cite{Oster_et_al2013} demonstrate how individuals at risk for Huntington's disease display optimism by deciding not to undergo genetic testing for early detection. This decision reflects a form of optimistic bias, where individuals may avoid information that could potentially challenge their desired outcomes or beliefs. Similarly, \cite{Engelman_et_al_2024} provides evidence that people engage in WT to alleviate anxiety about adverse future outcomes. \cite{orhun2021motivated} analyze beliefs about the risks of returning to work during a pandemic, and \cite{seaward2000optimism} investigate WT behavior concerning the repayment time for student loan debt.

\smallskip

In the context of markets,  \cite{Grubb_2015} and \cite{stone2018cognitive} provide theoretical and empirical evidence of consumer and firm over-optimism, respectively. \cite{Malmendier_Taylor_2015_second}  offer insights into CEOs' overconfidence, while \cite{Kent_Hirshleifer_2015} discuss WT's role in explaining investors' optimistic behavior in financial markets. Furthermore, \cite{lovallo2003delusions}  delve into the optimistic tendencies and overconfidence of business executives and entrepreneurs in their decision-making processes.\footnote{ \cite{Benabou_Tirole_2016} provide an in-depth summary of the evidence regarding optimism in economic decisions.}  Works such as \cite{taylor1991asymmetrical} and \cite{baumeister2001bad} study the tendency to focus on bad outcomes and ignore good outcomes. However, works in psychology have provided robust foundations and explanations for the prevalence of optimistic biases. In fact, \cite{korn2014depression} find that it is primarily depressed subjects who seem to follow objective probabilities, while most psychologically ``healthy'' individuals display some degree of optimism and biased updating. As such, similar mechanisms that push pessimists to ignore good outcomes likewise push optimists to downplay bad outcomes in their analysis. A recurrent theme throughout the literature is the fact that the relevance of optimism vs. pessimism is both individual-specific (\cite{hecht2013neural}) and context-specific (\cite{menon2009biases}).\footnote{With this in mind, we note that our analysis is easily extended to a pessimistic environment, and this may prove a fruitful direction for future research} 

\smallskip

In this paper, we develop a model of WT that considers both the benefits and costs associated with biased beliefs and optimistic behavior. Drawing on the framework established by \cite{bracha2012affective} and \cite{CaplinLeahy2019}, we study a two-stage model where a decision-maker (DM) confronts uncertainty about future events and makes choices involving actions and belief structures. By actively selecting their beliefs, the DM aims to maximize her subjective expected utility (SEU) for a given alternative, considering the cost of deviating from prior beliefs. To quantify this cost, we introduce the threshold beliefs distortion cost function, which penalizes deviations from prior beliefs in a binary fashion. Formally, our cost function is a convex (not strictly convex) non-smooth statistical distance  (in the sense of \cite{Csiszar1967})  between the prior and subjective beliefs.\footnote{\cite{BurghMelo2023} study a model similar to the one studied in this paper, but with a belief distortion cost which is assumed to be strictly convex and twice differentiable.}

\smallskip

We make multiple contributions to the literature on WT. First, our model uncovers a connection between WT behavior and the notion of superquantile optimization (\cite{Rockafellar2000OptimizationOC}, \cite{ROCKAFELLAR20021443}, and \cite{Rockafellar_Royset}). Intuitively, the superquantile of a random variable corresponds to the average of the top $(1-\alpha)100\%$ outcomes, where $\alpha$ is a parameter between $0$ and $1$. Furthermore,  this connection can be used to analyze median WT behavior, i.e., $\alpha=0.5$, the traditional expected utility case, $\alpha=0$, and extreme optimism with $\alpha=1$, a situation where the DM only looks at \emph{the highest} utility level of each alternative. Accordingly, we interpret $\alpha$ as the degree of \emph{optimism}.
\smallskip

Our analysis reveals that an optimistic DM focuses on the upper tail of utility outcomes associated with each alternative. Subsequently, the DM selects the option with the highest superquantile. As a result, an optimistic DM concentrates solely on favorable outcomes, which are those in the upper tail of the distribution of utility realizations for a given alternative. The critical driver behind this result is that optimism distorts the belief distribution by placing additional weight on the probability of good outcomes by reassigning weight from undesirable outcomes, a concept we refer to as censorship. Notably, we will demonstrate that this subjective probability assigns undesirable states probability zero.

\smallskip

This finding establishes a formal connection between WT and models of quantile-utility maximization, such as those found in \cite{Chambers_2009}, \cite{Manski_1988}, \cite{deCastro_Galvao_2019_ECMA,deCastro_Galvao_2021_ET}, and \cite{rostek2010quantile}. However, our approach differs from this literature in two key aspects. First, the connection between WT and quantile-utility maximization arises as a consequence of distorted beliefs due to optimism and not as a consequence of the primitives (or axioms) of the model. Second, our analysis demonstrates that an optimistic DM is not solely concerned with the utility associated with a specific quantile but also the conditional average utility related to the upper tail defined by the specific quantile. In other words, our analysis accounts for low-probability large-utility tail events, while the quantile approach does not account for this information. It is important to note that, as decision criteria, quantile preferences and our superquantile approach do not necessarily generate the same predictions.
\smallskip

In our second contribution, we leverage the connection between superquantile-utility maximization and WT behavior to shed light on DM's preference for skewness. Specifically, when the utility of different alternatives follows a Pareto distribution, we demonstrate that the optimal choice of a WT agent depends on the shape of the right tail or degree of positive skewness. The shape parameter, which determines the tail behavior, is crucial in determining the DM's optimal action.

\smallskip

Our third contribution is the development of an optimistic stochastic choice framework. Superquantiles have a monotonicity property that extends naturally to a notion of monotone stochastic choice. We achieve this by making use of results from \cite{Ballester_2018}. The monotonicity of stochastic choice probabilities in the degree of optimism provides testable implications for changes in DM's optimism levels.

\smallskip

Finally, we present an application of WT in a market entry decision. We explore how the shape of potential profit realization influences an optimistic entrepreneur's choice to enter a market. Moreover, noting that both risk aversion and optimism play a role in entrepreneurial endeavors (\cite{aastebro2014seeking}), we demonstrate a choice criterion that accounts for both optimism and risk aversion.

\subsection{Related Literature}\label{relatedlit}  Our paper is situated within several strands of literature. Primarily, it aligns with the literature on WT and motivated reasoning within economic decision-making, as discussed by  \cite{Benabou_Tirole_2016}. Notably, our work shares similarities with the studies by \cite{bracha2012affective}, \cite{CaplinLeahy2019}, and \cite{BurghMelo2023}. These papers explore decision-making models where a DM selects a probability distribution over states to maximize the difference between SEU and the cost of distorting beliefs. \cite{CaplinLeahy2019} focus on a cost function proportional to the Kullback-Leibler distance between subjective and prior beliefs, while \cite{BurghMelo2023} concentrate on strictly convex twice-differentiable cost functions, where the Kullback-Leibler divergence is a particular instance. \cite{bracha2012affective} focus on general convex and smooth cost functions.

\smallskip

The paper at hand introduces the threshold beliefs distortion cost function, distinct for not assuming strict convexity or smoothness as seen in \cite{bracha2012affective}, \cite{CaplinLeahy2019}, and \cite{BurghMelo2023}. We leverage our cost function to establish the equivalence between WT behavior and superquantile optimization. Another crucial distinction between these results and those at hand is a formal connection between WT behavior and a preference for positive skewness. Notably, neither the superquantile connection nor the preference for skewness result can be derived from the findings of \cite{bracha2012affective}, \cite{CaplinLeahy2019}, \cite{BurghMelo2023}. 
 
\smallskip

In a distinct line of work, the papers by \cite{mayraz2019priors} and \cite{kovach2020twisting} offer an axiomatic foundation for WT behavior. These papers contribute to understanding WT behavior through the lens of behavioral axioms. However, it is essential to highlight that neither of these papers incorporates the cost associated with belief distortion into their respective frameworks. \cite{bracha2012affective} presents an axiomatic foundation for WT behavior that incorporates this cost by amending the axiomatic characterization of variational preferences in \cite{maccheroni2006ambiguity}.
\smallskip

Secondly, our research contributes to the literature on optimal expectations, specifically referring to the work of \cite{BrunnermeierParker_20005}. Their study focuses on a dynamic model of belief choice, albeit without including the cost associated with maintaining optimistic beliefs. In contrast to our approach, the model of \cite{BrunnermeierParker_20005} explores the dynamics of belief choice in generating a preference for skewness in a portfolio allocation problem. It is crucial to highlight that our model's preference for skewness results does not imply their result, and vice versa. This distinction arises from fundamental differences in the nature and structure of the two models, emphasizing the unique contributions of each to the understanding of optimal expectations.\footnote{For an empirical test of optimal expectation models, we refer the reader to \cite{Oster_et_al2013}.}
\smallskip

Finally, our paper is also related to the literature on robustness in economic models. Specifically,  \cite{Hansen_Sargent_2001, Hansen_Sargent2008} introduce the idea of robust optimization to address the concern of model misspecification. They adopt a max-min approach, akin to multiple priors models as in \cite{GILBOA1989141} and \cite{maccheroni2006ambiguity}, to make decisions under ambiguity. While robustness and ambiguity models typically take a pessimistic approach to decision-making, our paper focuses on WT behavior and adopts an optimistic approach by studying a max-max problem. By doing so, our paper draws on the active and rapidly growing literature on distributionally robust optimization problems (\cite{Shapiro} and \cite{Kuhn_et_al_2019}). We leverage this mathematical framework to model WT behavior and derive its implications for economic decision-making. Thus, while our paper is related to the robustness literature, it differs in terms of its focus on WT, adopting an optimistic approach, and using distributionally robust optimization techniques. 
\smallskip

The remainder of the paper is structured as follows: in Section \ref{model}, we introduce the model and characterize optimal beliefs. Section \ref{section3} explores the connection between WT behavior and superquantiles and discusses the relationship and distinctions with the quantile utility maximization approach. Section \ref{s5} delves into the connection between WT behavior and positive skewness. Section \ref{concusions} provides the concluding remarks. 

\section{The Model}\label{model}

In this section, we develop a WT model that incorporates the benefits and costs of distorted beliefs. As the introduction section mentioned, we build upon  \cite{bracha2012affective} and \cite{CaplinLeahy2019}'s framework. In particular, we follow their idea that the DM   maximizes her current subjective expected utility, which incorporates utility from current experience (assumed zero) and utility from the DM's anticipated future realization. This relies on the view that an agent's subjective utility depends on beliefs regarding future outcomes.
\smallskip

Formally, we consider an environment where the DM is confronted with the task of selecting an action $a$ from a set $A=\{a_1,\ldots,a_n\}$ in the presence of uncertainty regarding a utility-relevant state $\omega \in \Omega$. We assume that $\Omega\subseteq \RR^n$ is a continuous set of states. Accordingly $\omega$ is a $n$-dimensional vector. The DM's utility function is defined as $u: A\times \Omega\longrightarrow \mathbb{R}$, which assigns a real-valued number to each pair consisting of an action and a state. The DM is endowed with an exogenous prior belief $F$, representing a probability distribution over $\Omega$. We assume that $F$ has a well-defined density denoted by $f$. The prior belief  $F$ can represent an objective likelihood of a state occurring, a data-driven probability based on similar (or previous) experiences, or the beliefs of an expert. For each $a\in A$, $\EE_F(u(a,\omega))$ denotes the (objective) expected utility. It is assumed that $\EE_{F}(|u(a,\omega)|)<\infty$ for all $a\in A$. 
\smallskip

The DM forms a subjective belief represented by the probability distribution $G$ over $\Omega$. Intuitively, $G$ can be interpreted as a distortion of $F$.\footnote{ As we shall see, in general the subjective belief $G$ will depend in the optimal action $a\in A$.} Accordingly, the subjective expected utility (SEU) of alternative $a \in A$ for the DM is given by $\mathbb{E}_{G}(u(a,\omega))$.  This expected payoff makes explicit that the DM uses her subjective beliefs $G$ to evaluate utility-maximizing actions.  
\smallskip

Following \cite{bracha2012affective} and \cite{CaplinLeahy2019}, deviating from $F$ induces a cost for the DM. This cost reflects the DM's \textit{taste for accuracy} and reflects a cognitive cost of distorting holding beliefs away from priors. These can reflect the internal cost of processing information (\cite{trimmer2016optimistic}) or future material costs of behavior under inaccurate beliefs (\cite{akerlof1982economic}). We introduce a novel belief distortion cost function, highlighting that large distortions from reality are highly costly. Formally, we model the cost of belief distortion $C_\alpha(G\|F)$ by leveraging the concept of $\phi$-divergence between the subjective belief $G$ and the prior $F$. For a given convex function $\phi$, \cite{Csiszar1967} defines the $\phi$ divergence as $C(G\|F)=\int_{\Omega}f(\omega)\phi\left(g(\omega)/f(\omega)\right)d\omega$. This cost $C(G\| F)$ captures a notion of \textit{statistical distance} between subjective and prior beliefs.\footnote{\cite{BurghMelo2023} study a WT model focusing on the class of strictly increasing, strictly convex, and smooth $\phi$-divergence functions. All of these properties are violated by the $\phi$ function defined in (\ref{Indicator_phi}). However, as we shall see, none of those assumptions are required in our analysis. }
\smallskip


Accordingly, given the cost $C_{\alpha}(G\|F)$, the WT agent chooses an optimal pair $(a^\star,G^\star)$ that maximizes:
\begin{equation}\label{wishful_thinking_problem}
  \max_{a\in A}\max_{G \in \mathcal{M}(F)}  \left\{\EE_G(u(a,\omega)) - \delta C_{\alpha}(G\|F)\right\}
\end{equation}
where $\mathcal{M}(F)$ is a set of absolutely continuous distributions with respect to $F$.
\smallskip 

The infinite-dimensional problem (\ref{wishful_thinking_problem}) presents a framework in which the agent simultaneously chooses both an action and a belief structure. From a behavioral standpoint, this framework considers the subjective beliefs to be contingent on the chosen action. In other words, the DM assigns a specific belief structure to each possible action, allowing for the possibility of holding seemingly contradictory beliefs. This notion of beliefs being action-contingent resembles the concept of \emph{cognitive dissonance}, as discussed in \cite{akerlof1982economic} and relates to situations where agents may hold contradictory beliefs.
\smallskip

It is worth pointing out that our way of modeling  WT behavior is closely connected to motivated reasoning, as discussed in studies such as \cite{Kunda1990TheCF} and \cite{Benabou_Tirole_2016}. According to this theory, when choosing their optimal beliefs, a motivated DM is driven by the SEU $\EE_G(u(a,\omega))$, representing the anticipated utility associated with different actions and outcomes. However, in addition to the utility, a motivated DM also considers the importance of accuracy.  This latter aspect is captured by our cost function $ C_\alpha(G\|F)$. Thus, our model incorporates the motivational aspect of maximizing SEU and considering accuracy in belief choice.
\smallskip

Now, we turn our attention to modeling the cost $C_\alpha(G\|F)$. In this regard, we specifically focus on the following $\phi$-divergence function:

\smallskip

\begin{equation}\label{Indicator_phi}
   \phi(t)=\begin{cases}
	0, & 0\leq t\leq \frac{1}{1-\alpha}\\
	+\infty & \text{otherwise},
\end{cases}
\end{equation}
with $\alpha\in (0,1).$
\smallskip

Using $\phi$-function (\ref{Indicator_phi}), we define $C_\alpha(G\|F)$ as the corresponding $\phi$ divergence between G and F. This yields a cost of distorting beliefs about state $\omega$ such that the cost is zero whenever the ratio ${g(\omega)/f(\omega)}$ is between $0$ and \\$(1-\alpha)^{-1}$. Conversely, if ${g(\omega)/f(\omega)}$ lies outside this range, then $C_\alpha(G\|F)=+\infty$. Formally, (\ref{Indicator_phi}) induces the cost of belief distortion:

\smallskip
 \begin{equation}\label{Cost_phi}
   C_\alpha(G\| F)=\begin{cases}
	0, & 0\leq \frac{g(\omega)}{f(\omega)}\leq \frac{1}{1-\alpha},  \; \forall \; \omega \in \Omega\\
	+\infty & \text{otherwise}
\end{cases}
\end{equation}
with $\alpha\in (0,1).$

\smallskip

The cost function (\ref{Cost_phi}) models a situation where the DM is allowed to pick any probability distribution $G$ as long as the likelihood ratio $g(\omega)/f(\omega)$ is in the range of $0$ and $(1-\alpha)^{-1}$. Similar to \cite{akerlof1982economic}, our cost function captures the role of cognitive dissonance in the decision-making process.

\smallskip

Given   the structure of (\ref{Cost_phi}), we refer to $C_\alpha(G\|F)$ as the \emph{threshold beliefs distortion cost} induced by $\phi$-divergence function (\ref{Indicator_phi}). We note that the parameter $\alpha$ is key in  defining $\phi$ and $C_\alpha(G\| F)$, respectively. In fact, as we will see, $\alpha$ captures the DM's degree of optimism. 
 \smallskip
 
 The simplicity  of the cost (\ref{Cost_phi}) allows us to express the inner maximization problem in (\ref{wishful_thinking_problem}) in an alternative and more  tractable way.

\begin{proposition}\label{dualprob} Let $V_\alpha(a)\triangleq\max_{G\in\mathcal{M}(F)}\left\{\mathbb{E}_{G}(u(a,\omega))- C_\alpha(G\|F) \right\}$ for all $a\in A$. Then, the following holds:
\begin{equation}\label{CVaR}
V_\alpha(a)=\min_{\lambda_a\in\RR}\left\{\lambda_a+{1\over 1-\alpha}\EE_F(\max\{u(a,\omega)-\lambda_a,0\})\right\}
\end{equation}
\end{proposition}
\proof All proofs are collected in the appendix \eproof

\medskip

The previous proposition establishes that the infinite-dimensional problem of finding an optimal belief $G$ can be equivalently expressed  as a  simpler one-dimensional optimization program.  This equivalence dramatically simplifies the analysis of (\ref{wishful_thinking_problem}) because the problem (\ref{wishful_thinking_problem}) can be expressed only in terms of the prior distribution $F$. More importantly, the equivalence (\ref{CVaR})  allows us to express our WT  model in terms of a superquantile optimization problem (\cite{Rockafellar2000OptimizationOC}, \cite{ROCKAFELLAR20021443}, and \cite{Rockafellar_Royset}).  We will exploit this fact to show how a WT agent will look at the upper quantiles of the outcome distribution censoring other sections of the distribution associated with less desirable outcomes. In addition, we will use (\ref{CVaR}) to characterize optimistic decision-making in terms of risk and skewness.
\smallskip

Finally, we note that the Proposition \ref{dualprob} follows from  a direct application of Lemma 1 in \cite{BurghMelo2023}.  Furthermore, leveraging their Proposition 2, we can interpret $V_\alpha$  as a risk measure. From a behavioral standpoint, this fact implies that WT agents behave as risk seekers.\footnote{In particular, we note that in the portfolio optimization literature, $V_\alpha$ is known as conditional value-at-risk (CVaR) (\cite{Rockafellar2000OptimizationOC}).} 

\smallskip

\section{Optimal choice and superquantile maximization}\label{section3}
In this section, we show that WT behavior corresponds to superquantile-utility maximization and preference for skewness. 
\smallskip

Recall that the prior distribution over states is given by $F$ with density function  $f$. For each $a\in A$, we define a cumulative distribution function of the utility $u(a,\omega)$ it yields. Define the distribution corresponding with action $a$ as:

\begin{equation*}\label{Induced_Distribution}
 F_a(z)\triangleq \PP(\{\omega\in \Omega: u(a,\omega)\leq z\}).    
\end{equation*}

\medskip

 We assume that $F_a(z)$ is strictly increasing and continuous on the interior of its domain. We make this assumption for the sake of simplicity.\footnote{Our results naturally extend to the case where $F_a(z)$ is not strictly increasing and to the case where atoms are allowed.} 
 We now introduce the familiar notion of quantile.

 \smallskip
 
\begin{definition}\label{VaR_Definition} For $a\in A$, the $\alpha$-quantile of the random variable $u(a,\omega)$ is given by: 
\begin{equation}\label{VaR_Definition_Equation}
 Q_\alpha(a)\triangleq \min \{z\in \RR \mid F_a(z) \geqslant \alpha\}, \alpha \in(0,1).
 \end{equation}
\end{definition}

\smallskip

In the preceding definition, note that the assumption stipulating $F_a(z)$ as strictly increasing and continuous on the interior of its domain implies that the quantile $Q_\alpha(a)$ is uniquely defined.
\smallskip

Next, we introduce the concept of superquantile as follows:
\begin{definition}[\cite{Rockafellar_Royset}]\label{CvAR_Definition} For  $a\in A$,   the $\alpha$-superquantile of the random variable $u(a,\omega)$ is given by:
\begin{equation}\label{CVAR_Equation_Def}
\bar{Q}_{\alpha}(a)\triangleq Q_\alpha(a)+{1\over 1-\alpha}\EE_{F}\left(\max\{u(a,\omega)-Q_\alpha(a),0\}\right)\quad \alpha\in (0,1),
\end{equation}
where $Q_{\alpha}(a)$ is the $\alpha$-quantile  defined in (\ref{VaR_Definition_Equation}).
\end{definition}

Intuitively, the notion of $\alpha$-superquantile  corresponds to the average of outcomes beyond the specified $\alpha$-quantile $Q_{\alpha}(a)$. As such, we also can make use of equivalent definitions of the superquantile throughout the paper:

$$\bar{Q}_{\alpha}(a) = \frac{1}{1 - \alpha}\int_\alpha^1Q_\theta(a) \; d\theta = \frac{1}{1 - \alpha}\int_{\{\omega\in \Omega: u(a,\omega)\geq Q_\alpha(a)\}} f(\omega)u(a,\omega) \; d\omega.$$
\smallskip

With Definitions \ref{VaR_Definition} and \ref{CvAR_Definition} in place,  we are ready to establish the main result of this section.

\smallskip

\begin{proposition}\label{WT_CvAR} Consider the WT problem (\ref{wishful_thinking_problem}) with $C_\alpha(G\| F)$ given by threshold beliefs distortion cost function (\ref{Cost_phi}). Then, for $\alpha\in (0,1)$ an optimal solution pair $(a^\star,G^\star_{a^\star})$ satisfies the following:
\begin{itemize}
\item[(i)] The optimal $\lambda_{a^\star}^\star$ in (\ref{CVaR}) satisfies $\lambda_{a^\star}^\star=Q_{\alpha}(a^\star).$
\item[(ii)]$V_\alpha(a^\star)=\bar{Q}_{\alpha}(a^\star)$
\item[(iii)] $a^\star$  satisfies:
\begin{equation}\label{Expected_Shortfall_WT}
a^\star=\arg\max_{a\in A}\left\{\EE_{F}(u(a,\omega)|u(a,\omega)\geq Q_\alpha(a))\right\},\nonumber
\end{equation}
where $\bar{Q}_{\alpha}(a)=\EE_{F}(u(a,\omega)|u(a,\omega)\geq Q_\alpha(a))
\}$ for all $a\in A.$
\item[(iv)] The  distorted  belief $G_\alpha(a^\star)$ corresponds to:
$$G_\alpha(a^\star,\omega)\triangleq \begin{cases}\frac{F_{a^\star}(\omega)-\alpha}{1-\alpha} & \text { if } u(a^\star,\omega) \geq Q_{\alpha}(a^\star) \\ 0 & \text { if } u(a^\star,\omega)<Q_{\alpha}(a^\star) .\end{cases}$$

\end{itemize}
\end{proposition}

\medskip

The previous proposition fully characterizes WT behavior in terms of $u(a, \omega)$, $Q_\alpha(a)$, and $\bar{Q}_\alpha(a)$. Parts (i) and (ii) establish that the optimal $\lambda^\star_{a^\star}$ corresponds to the $\alpha$-quantile, and $V_{\alpha}(a)$ is the $\alpha$-superquantile, respectively. These two facts allow us to formalize the observation that, in our model, WT behavior is tailored around a specific segment of the prior distribution $F$. This behavioral implication is captured in part (iii), which establishes that the WT problem boils down to selecting the action with the highest $\alpha$-superquantile, as measured by $\EE_F(u(a, \omega)|u(a, \omega)\geq Q_\alpha(a))$. This implies that a WT agent focuses exclusively on the upper quantiles of the distribution associated with each random variable $u(a, \omega)$. Thus the DM censors the bottom portion of the distribution of $u(a, \omega)$. Consequently, the DM faces a cognitive bias that leads her to rely on a \emph{truncated distribution}, effectively disregarding outcomes below the $\alpha$-quantile $Q_{\alpha}(a^\star)$. As a result, WT behavior implies that the DM concentrates exclusively on the tail segment and chooses the action with the highest \emph{expected upper-tail utility}. Truncated subjective beliefs are noted as an explanation for the entrepreneurial puzzle by \cite{deligonul2008entrepreneuring}.\footnote{Here, the entrepreneurial puzzle refers to observations of entrepreneurial activity in situations where risk-return levels are significantly lower than those of private and public equity indexes}

\smallskip

To our knowledge, Proposition \ref{WT_CvAR}(iv) constitutes the first formalization of this type of censored belief cognitive bias in economic behavior.\footnote{It is worth noting that in a different setting, \cite{Benabou2013} examines a model where a decision-maker censors bad states, inducing optimistic behavior. Our characterization differs from his in that we consider both the benefits and costs of holding optimistic beliefs.} Moreover, this characterization formalizes the connection between WT behavior and the notion of a preference for skewness. This preference becomes evident when examining the expression in part (iii), which captures the behavior of a WT agent favoring alternatives with high utility levels even if they have a low probability of occurrence. The DM overemphasizes these high payoff outcomes, focusing on expected upper-tail utility, while parameter $\alpha$ determines how much extra weight is placed on the tails. In other words, Proposition \ref{WT_CvAR}(iii)-(iv) elucidates the decision-maker's inclination towards skewness by explicitly addressing considerations regarding tail performance comparisons. In Section \ref{s5}, we further discuss this result, providing closed-form expressions that assist in understanding WT behavior in terms of tail distributions and a preference for skewness.

\smallskip

It is worth pointing out Proposition \ref{WT_CvAR} establishes that WT behavior is intimately related to \emph{quantile-utility maximization} (\cite{Chambers_2009}, \cite{deCastro_Galvao_2019_ECMA, deCastro_Galvao_2021_ET}, \cite{rostek2010quantile}, and \cite{Manski_1988}). However, unlike quantile preference models, the result in Proposition \ref{WT_CvAR} indicates that an optimistic DM considers not only the utility \emph{at the quantile level} but also the average \emph{expected tail utility} associated with it. More importantly, the characterization in Proposition \ref{WT_CvAR} makes transparent the fact that WT behavior leads a DM to consider a distorted belief distribution over the set of states. Furthermore, quantiles and superquantiles models do not necessarily agree on the prescribed optimal solutions. The following example, based on \cite{Royset2023riskadaptive} illustrates this point.

\smallskip

\begin{example}\label{Example_Comparing_QP_SQP} Let $A=\{-1,1\}$.  Assume that $\omega $ follows a  triangular distribution in $[0,2]$ with mode at $0$ The utility associated with each alternative $a$ is given by $u(a,\omega)=(\omega-2 / 3) a-1/3$. Using the distributional assumption on
it is easy to see that $\EE_{F}(u(1,\omega))=\EE_{F}(u(-1,\omega))=-1/3$. Now, if we use a quantile utility-maximization approach, we get $Q_{0.8}(1)=0.106$ and $Q_{0.8}(-1)=0.122$. According to this approach, the alternative $a=-1$ is selected. Now, we can compute $\bar{Q}_{0.8}(1)=0.404$ and  $\bar{Q}_{0.8}(-1)= 0.230$. Using the superquantile criterion, the decision is reversed, implying that a DM will choose $a=1$. The reason for this reversal comes from the fact that the 
quantile approach ignores  the magnitude of outcomes above $Q_{0.8}(1)$ and $Q_{0.8}(-1)$ and thus fails to reflect the difference of the average  utility levels beyond the 0.80-quantile.\eproof
\end{example}

\smallskip

From a behavioral perspective, Proposition \ref{WT_CvAR} underscores the significance of $\alpha$ in reflecting the DM's degree of optimism. Specifically, $\alpha$ represents the weight assigned by the DM to events in the upper tail of the distribution. More critically, $\alpha$ can be interpreted as the degree of selectivity or “censorship” exercised by the DM when evaluating and choosing between alternatives. The following result formalizes the relationship between $\alpha$ and the behavior of $V_\alpha(a)$.

\begin{corollary}\label{WT_CvAR_Corollary} Consider the WT problem (\ref{wishful_thinking_problem})  with $C_\alpha(G\| F)$ given by threshold beliefs distortion cost function (\ref{Cost_phi}). Then for all $a\in A$:
\begin{itemize}
\item[(i)] $V_{\alpha}(a)$ is continuous in $\alpha$.
\item[(ii)]
 $\frac{\partial V_\alpha(a)}{\partial \alpha}=(1-\alpha)^{-1}\EE_{F}(\max\{u(a,\omega)-\lambda_a^*,0\})$.
 \item[(iii)]
$V_\alpha(a)$ is monotone non-decreasing in $\alpha$. In particular, for $\alpha_1\leq \alpha_2$ we get $V_{\alpha_1}(a)\leq V_{\alpha_2}(a)$.
\item[(iv)] When $\alpha\longrightarrow 0$ we get
$$V_{\alpha}(a)\longrightarrow\EE_{F}(u(a,\omega)).$$
\item[(v)]When $\alpha\longrightarrow 1$ we get:
$$V_{\alpha}(a)\longrightarrow\sup_{\omega\in \Omega}u(a,\omega),$$
where $\sup_{\omega\in \Omega}u(a,\omega)$ refers to the essential supremum of the random variable $u(a,\omega)$.
\end{itemize}
\end{corollary}

Part (i) establishes the distorted utility $V_\alpha(a)$  varies continuously  with respect to the degree of optimism $\alpha$. (ii) provides a simple formula that characterizes how the $V_\alpha(a)$ responds to changes on $\alpha.$ This result is complemented by (iii). Finally, parts (iv) and (v) formalize s the limiting cases for $\alpha\longrightarrow 0$ and $\alpha\longrightarrow 1$, respectively.
  
\smallskip

We close this section pointing out that our WT can be shown to be behaviorally equivalent to EU maximization. Specifically, Proposition 3 in the work by \cite{BurghMelo2023} allows us to assert that WT behavior can always be interpreted as the outcome of EU maximization under a distorted utility function. An important implication of this finding is that, based on choice data alone, we cannot distinguish between decisions made under the WT model and those made under the traditional EU model.

\section{Optimistic discrete choice}\label{s5} 

In this section, we develop a simple discrete choice model with WT behavior. We consider a choice set $A=\{a_1,\ldots,a_n\}$ and assume that the outcome space $\Omega=\mathbb{R}^n$. In this context, each $\omega$ corresponds to an $n$-dimensional vector $\omega=(\omega_{a_1},\ldots,\omega_{a_n})$, where $\omega_a$ represents the realization associated with alternative $a$. The prior belief is represented by a distribution $F$ that is fully over $\Omega$. Let $F_a(\omega_a)$ denote the marginal distribution governing realizations of $\omega_a$. Without loss of generality, for all $a\in A$  we assume that $\mathbb{E}_{F_a}(\omega_a)=0$. With a slight abuse of notation, we denote the $\alpha$-quantile of random variable $\omega_a$ as $Q_\alpha(\omega_a)$ and the corresponding superquantile $\bar{Q}_{\alpha}(\omega_a)=\EE_{F_a}(\omega_a|\omega_a\geq Q_\alpha(\omega_a))$.
\smallskip

The utility associated with alternative $a$ is defined as \footnote{The additive structure is assumed for simplicity. However, our analysis  and results hold for the general case where  the utility associated with  each alternative $a\in A$ takes the general non-additive form $u(a,\omega_a)$.}
\smallskip
\begin{equation}\label{Optimistic_RUM_eq}
u(a,\omega)=u(a)+\omega_a.
\end{equation}

The DM has optimistic beliefs with respect to the realizations of $\omega$. 
In particular, the DM solves the optimistic optimization problem (\ref{wishful_thinking_problem}). Accordingly, from Proposition \ref{WT_CvAR}(i),  we know that for each alternative $a\in A$, the $\alpha$-quantile corresponds to  

$$\lambda_{a}^*=Q_\alpha(a) = u(a)+Q_{\alpha}(\omega_a).$$

The previous equality follows from the translation invariance of quantiles, a feature shared by superquantiles.  In particular, it follows that $\bar{Q}_\alpha(a)=u(a)+\bar{Q}_\alpha(\omega_a)$. Then, from Proposition  \ref{WT_CvAR}, we  know that  a  WT agent's optimal action is the solution to the following discrete choice problem:
\begin{equation}\label{WT_RUM_sol}
\max_{a\in A}\{u(a)+\bar{Q}_{\alpha}(\omega_a)\}.
\end{equation}

In the latter expression, the WT agent considers both the deterministic utility $u(a)$ and the average upper tail utility associated with  $\omega_a$ as measured by $\bar{Q}_\alpha(\omega_a)$. Specifically, in equation (\ref{WT_RUM_sol}), the $\alpha$-superquantile $\bar{Q}_{\alpha}(\omega)$ reflects the level of optimism associated with alternative $a$. From a behavioral perspective, this implies that the WT agent \emph{overestimates} the utility of each alternative.
\smallskip

This upward bias arises because $\bar{Q}_{\alpha}(\omega_a)\geq \mathbb{E}_{F_a}(\omega_a)=0$, indicating that the WT agent assigns higher perceived utility to the alternative $a$ by focusing on the positive upper tail outcomes of $\omega_a$. By selectively considering the upper tail realizations and censoring the lower tail or average outcomes, the WT agent exhibits a biased perception of the actual utility of each alternative. This bias reflects the optimistic belief that the upper tail outcomes will occur more frequently or have a more significant impact than they do in reality.
\smallskip

It is important to note that the additive utility structure presented in equation (\ref{Optimistic_RUM}) bears similarities to the well-known additive random utility model (ARUM) (\cite{McFadden1978, mcf1}). However, there are two significant differences between the two approaches. First, the ARUM framework does not explicitly incorporate optimistic behavior. In contrast, our model allows for WT by considering each alternative's average upper tail utility values. This introduces an element of optimism into the decision-making process, leading to potentially different choice outcomes compared to the ARUM. Second, the ARUM approach provides an optimal stochastic choice rule describing the probabilities of each alternative $a \in A$. In contrast, our discrete choice model provides a \emph{deterministic} choice rule specifying how the DM selects a particular alternative based on WT behavior. The expression (\ref{WT_RUM_sol}) serves as a prescription for decision-making under wishful thinking.
\smallskip

In Section \ref{Optimistic_RUM}, we discuss the implications of incorporating censored beliefs and optimistic behavior into a stochastic choice framework. Specifically, we demonstrate how our model provides a more nuanced understanding of decision-making by accounting for the influence of optimistic beliefs in deriving an optimal stochastic choice rule consistent with ARUM.

\begin{example}\label{Example2}To see how the previous model provides new insights let us assume that for each $a\in A$, the random variable $\omega_a$ follows a normal distribution with mean zero and variance $\sigma_a^2$. Using  the results in \cite{Uryasev_et_al_2018}, the problem (\ref{WT_RUM_sol}) can be expressed as:
$$\max_{a \in A}\left\{u(a)+\sigma_a{f(Q_\alpha(\omega_a))\over 1-\alpha}\right\}$$
Thus, when the $\omega_a$s are normally distributed, the distorted utilities can be interpreted as a mean-variance term. Specifically, for each alternative $a$, the distorted utility increases with the standard deviation $\sigma_a$. Consequently, in this environment, the WT agent assigns value to the risk associated with each alternative. This example highlights that a WT agent, prefers taking more risks than a traditional rational DM.

\end{example}

\smallskip

\subsection{Skewed discrete choice}\label{skewedsection} Our next goal is to understand the  factors shaping $\bar{Q}_\alpha(a)$ and consequently influencing the choice process. To this end,  we focus on the Pareto distribution family. The Pareto distribution is characterized by two parameters: $\bar{\omega}_a \in \mathbb{R}$ and $\beta_a > 0$. The Pareto density function can be expressed as follows:

$$F(\omega_a) = 1 - (\frac{\bar{\omega}_a}{\omega_a})^{\beta_a},$$
where $\omega_a \in [\bar{\omega}_a,\infty)$ is a scale parameter, and $\beta_a > 0$ is a shape parameter.

\smallskip

The shape parameter is of particular interest to us as it captures the thinness/thickness of the tail. It is important to note that for $\beta_a \leq 1$, $\EE(\omega_a) = \infty$, which implies the superquantile is not well defined ($\bar{Q}_\alpha(\omega_a) = \infty$). As such, we focus on the case where $\beta_a > 1$. This environment provides a closed-form expression for the superquantile  $\bar{Q}_\alpha(\omega_a)$, allowing us to uncover the direct role of the quantile in WT decision-making. Furthermore, the Pareto distribution yields an environment with positive skewness in the utility, allowing for insights into the value of the shape of tails. To analyze this case, we compute the value associated with the quantile $Q_{\alpha}(\omega_a)$. Making use of the expression for the density $F(\omega_a)$, we see:

$$Q_\alpha(\omega_a) = \lambda_a^* = \bar{\omega}_a(1-\alpha)^{-1/\beta_a}$$

Direct computation yields the superquantile:

$$\bar{Q}_{\alpha}(\omega_a)=(1 - \frac{1}{\beta_a})Q_\alpha(\omega_a)$$

\smallskip

The following proposition formalizes the previous analysis
\begin{proposition}\label{Prop_skewed_DC} Consider a WT problem with the additive payoff structure (\ref{Optimistic_RUM}). Assume that $\omega_a$ follows a Pareto distribution with parameters $\bar{\omega}_a \in \mathbb{R}$ and $\beta_a >1$ for each $a\in A$. Then the WT agent solves the following problem
\begin{equation}\label{WT_SOL_GPD}    
\max_{a\in A}\left\{u(a)+ (1 - \frac{1}{\beta_a})Q_\alpha(\omega_a)\right\}.
\end{equation}
\end{proposition}
\smallskip

This expression reveals that an optimistic DM will modify her utilities by incorporating a term that depends on the quantile $Q_{\alpha}(\omega_a)$ and the shape. Note that a higher $\beta_a$ value indicates a thin long tail, meaning fewer extreme values. This raises the term $(1-\frac{1}{\beta_a})$, reflecting the ability of the WT DM to overweigh these limited extreme value outcomes. This modification allows the DM to capture the effects of WT and tailor her preferences based on the specific characteristics of each alternative's positive skewness.\footnote{In the online appendix, we discuss several other distributions that yield closed-form formulas for $\EE_{Q_a}(\omega_a|\omega_a\geq Q_{a}^{-1}(\alpha))$, including the logistic distribution, GPD, and GEV.}

\subsection{Optimistic stochastic choice}\label{Optimistic_RUM}
We now discuss how our model can be embedded within the framework of stochastic choice. Specifically, we consider the case where the DM faces a binary choice set $A=\{a_1, a_2\}$. For simplicity, assume that the utility associated with $a_1$ is state-contingent, given by $u(a_1, \omega)=u(a_1)+\omega$, while the utility associated with $a_2$ is deterministic, given by $u(a_2, \omega)=u(a_2)$, where $u(a_1)<u(a_2)$. According to our censored belief model, the DM distorts the expected utility of $a_1$ by evaluating the $\alpha$-superquantile. When comparing $a_1$ and $a_2$, the DM analyzes the difference between $V_\alpha(a_1)=u(a_1)+\EE(\omega_{a_1} \mid \omega_{a_1} \geq Q_\alpha(a_1))$ and $V_\alpha(a_2)=u(a_2)$ for all $\alpha \in[0,1)$.

\smallskip

Following the literature on random utility and risk (e.g., \cite{Ballester_2018}), we consider an environment where the valuation of an action $a \in A$ is determined by the additive combination of $V_a(a)$ and a random, i.i.d., unobserved term $\epsilon_a$ that follows a continuous cumulative distribution $H$. Accordingly, we define the probability of selecting action $a_1$ over $a_2$ as
$$
\rho^{RUM}_{\alpha}\left(a_1,a_2\right)=\PP(V_{\alpha}(a_1)+\epsilon_{a_1} \geq V_\alpha(a_2)+\epsilon_{a_2}).
$$

\smallskip

To specify $\rho^{RUM}_{\alpha}\left(a_1,a_2\right)$, we adopt the widely used Gumbel (extreme value type 1) distribution for the unobserved terms. This assumption yields the popular Luce model (i.e., logit model). A key feature of the Luce model is that it provides closed-form expressions for the choice probabilities $\rho^{Luce}_{\alpha}\left(a_1,a_2\right)$ and $\rho^{Luce}_{\alpha}\left(a_2,a_1\right)$, respectively. Specifically, the Luce model yields the following stochastic choice rule:
\begin{equation}\label{MNL_Opt}
\rho^{Luce}_{\alpha}\left(a_1,a_2\right)={e^{V_\alpha(a_1)}\over e^{V_\alpha(a_1)}+e^{u(a_2)}}
\end{equation}
where $\rho^{Luce}_{\alpha}\left(a_2,a_1\right)=1-\rho^{Luce}_{\alpha}\left(a_1,a_2\right)$.
\smallskip

To illustrate the role of censored beliefs in shaping the stochastic choice rule (\ref{MNL_Opt}), consider that $\omega_{a_1}$ follows a standard normal distribution. From Example \ref{Example2}, we know that $V_\alpha(a_1)=u(a_1)+\frac{f(Q_\alpha(\omega_{a_1}))}{1-\alpha}$. Furthermore, from Corollary \ref{WT_CvAR_Corollary}, we know that as $\alpha \rightarrow 0$, we recover $V_0(a_1)=u(a_1)$. By assumption, $u(a_1)<u(a_2)$, so $\rho^{Luce}_0(a_1,a_2)<\rho^{Luce}_{0}(a_2, a_1)$. Thus, when the DM does not censor outcomes, the choice probability of the riskless option is higher.
\smallskip

\begin{figure}[h!]
    \title{\textbf{Stochastic Choice and Degree of Optimism}}
    \centering
    \includegraphics[width=10cm]{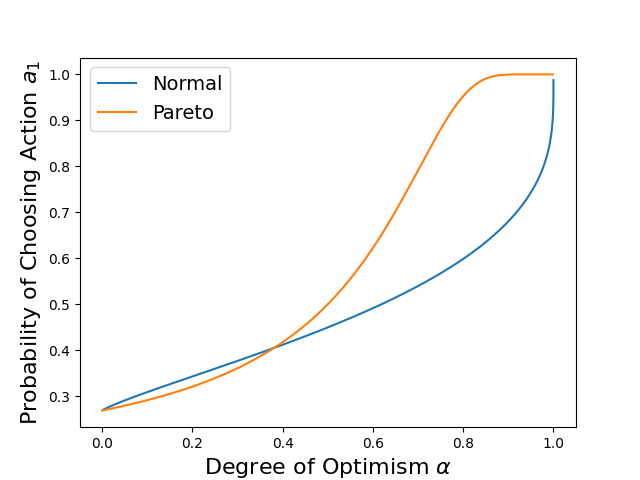}

    \medskip
    \caption{Choice probabilities $\rho _\alpha^{Luce}(a_1,a_2)$}
    \caption*{\tiny{This figure demonstrates how the level of optimism shapes stochastic choice behavior, where $\EE_F(u(a_1,\omega)) = 0$ and $u(a_2) = 1$}}
    \label{timingpayoffs}
\end{figure}

As $\alpha$ increases, the value of $V_\alpha(a_1)$ also increases (Corollary \ref{WT_CvAR_Corollary}(iii)), leading to an increase in $\rho^{Luce}_\alpha(a_1, a_2)$. More importantly, it is not difficult to find a threshold $\hat{\alpha}$ such that $V_{\hat{\alpha}}(a_1)>u(a_2)$, implying that the stochastic choice behavior reverses to $\rho^{Luce}_\alpha(a_1,a_2)>\rho^{Luce}_\alpha(a_2,a_1)$, and furthermore $\rho^{Luce}_\alpha(a_1,a_2) \rightarrow 1$ as $\alpha \rightarrow 1$.  Put formally, 
the difference $V_\alpha(a_1)-u(a_2)$ is \emph{monotone} increasing in $\alpha$.\footnote{The stochastic choice pattern described extends beyond the normal distribution case. For example, $\omega_{a_1}$ could follow a Pareto distribution, as in section \ref{skewedsection} or any of the distributions in the online appendix.} Figure \ref{timingpayoffs} demonstrates this monotonicity of stochastic choice behavior for the case of the normal and Pareto distributions, respectively.

\smallskip

The previous monotonicity pattern can be extended to the class of additive RUMs, where the discrete choice sets may contain more than two alternatives. Accordingly, let $\rho^{RUM}_\alpha$ denote the stochastic choice rule (choice probabilities) induced by an additive RUM,  where the DM evaluates the utilities $V_\alpha(a)+\epsilon_a$ for all $a\in A$, $\alpha\in [0,1)$, and $\epsilon_a$ that is a continuous random variable. Next, we adapt  \cite{Ballester_2018}'s monotonicity notion. Formally, for a pair of actions $(a,a^\prime)\in A$, let $\rho^{RUM}_\alpha(a,a^\prime)$ denote the probability of preferring $a$ over $a^\prime$ for all $\alpha\in[0,1)$. We say that $\rho^{RUM}$ is monotonically increasing for a pair of actions $(a,a^\prime)$ when $\rho^{RUM}(a,a^\prime)$ increases in $\alpha$ for all $\alpha \in (0,1)$. We can now demonstrate a tight condition for the monotonicity of $\rho^{RUM}$.
\smallskip
\begin{proposition}\label{Monotone_RUM_Opt} $\rho^{RUM}$ is monotonically increasing for the pair of actions $(a,a^\prime)$ if and only if for all $\alpha\in(0,1)$ the following inequality holds 
\begin{equation}\label{Mon_Inequality}
    \EE_{F}(\max\{ u(a,\omega)-Q_\alpha(a),0\})\geq \EE_{F}(\max\{u(a^\prime,\omega)-Q_\alpha(a^\prime),0\}), .
\end{equation}
 
\end{proposition}

Multiple remarks are in order. First, Proposition \ref{Monotone_RUM_Opt} builds off of results from Proposition 1 of \cite{Ballester_2018}. Our result establishes that when embedded into a RUM, our censored beliefs model can yield a monotone stochastic choice rule. This result provides a simple, testable implication of our optimistic RUM. Namely, policies targeted at increasing optimism may be tested directly through stochastic choice data. Second, we note that the degree of optimism,  $\alpha$, can be treated as a random parameter representing different “optimistic attitudes” within a population of DMs. This approach aligns with the random preference model discussed in \cite{Ballester_2018} in the context of risk aversion in stochastic choice models. A full exploration of this extension is left for future research.
\smallskip

Finally, we note that the monotonicity condition (\ref{Mon_Inequality}) does not rely on the additively separable specification (\ref{Optimistic_RUM_eq}). The condition is derived for a general case. For additive separable utility, condition (\ref{Mon_Inequality}) simplifies to:
\begin{equation*}\label{Mon_Inequality_Add}
    \EE_{F_a}(\max\{ u(a)-Q_\alpha(\omega_a),0\})\geq \EE_{F_{a^\prime}}(\max\{u(a^\prime)-Q_\alpha(\omega_{a^\prime}),0\}), 
\end{equation*}
for all $\alpha \in (0,1)$.

\subsection{Market entry decisions}\label{Entry_Market_model} We now apply the previous analysis to study a variant of the entry decision problem studied by \cite{BRESNAHAN199157}. Suppose a firm seeks to decide whether to enter a particular market. The action set is $A=\{enter, out\}$, where $enter$ represents the choice of entering the market, while $out$ represents staying out of the market. The firm faces uncertainty about this market's demand/competition captured by additive profit shock $\pi$. Staying out of the market yields profit $u(\text{out},\pi) = 0$, independent of the profit shock. Entering the market requires entry cost $k > 0$, so the profit associated with entry is given by $u(\text{enter},\pi) = \pi - k$. We assume that the profit shock $\pi$ follows a continuous distribution $F$ with  $\EE_F(\pi)=0$.
\smallskip

In this setting, a firm that seeks to maximize its expected profit would never choose to enter the market. This is because $\EE_F(u(out,\pi)) = 0 > -k = \EE(\pi - k) > \EE_F(u(enter,\pi))$. However, by applying the insights from Proposition \ref{WT_CvAR}, we see that an optimistic firm may choose to enter. Concretely, the firm will enter the market under specific parametric configurations, driving optimistic beliefs. To demonstrate, consider that choice is driven by optimistic beliefs as specified in Problem (\ref{wishful_thinking_problem}). For action $enter$, the optimal multiplier $\lambda_{Enter}^\star$ is given by $\lambda_{Enter}^\star = Q_{\alpha}(Enter) = Q_{\alpha}(\pi) - k$. Therefore, the firm chooses to enter if and only if the following condition holds:
$$\bar{Q}_{\alpha}(\pi)>k,$$
where $\bar{Q}_{\alpha}(\pi) =\EE_F(\pi|\pi\geq Q_{\alpha}(\pi)).$
\smallskip

Therefore, while the true mean of the profit shock distribution is zero, an optimistic disposition can lead it to instead rely on its \emph{subjective}  mean, captured by $\bar{Q}_{\alpha}(\pi)$. Furthermore, by defining $\hat{\alpha}$ such that $\bar{Q}_{\hat{\alpha}}(\pi)=k$, we note that the firm enters the market for all $\alpha \geq \hat{\alpha}$. The fact that only sufficiently optimistic firms choose to enter this market highlights the impact of WT on market entry, and aligns well with the well-documented connection between optimism and entrepreneurial endeavors (\cite{seaward2000optimism}).

\smallskip

Using the result in the Proposition \ref{Prop_skewed_DC}, we gain economic insights about how the skewness of distribution $F$ determines the firm's decision. Let us consider the case where the profit shock $\pi$ follows a Pareto distribution with density $F$ and shape parameter $\beta$. Using Proposition (\ref{Prop_skewed_DC}), we see that the WT firm will choose to enter the market if and only if the following condition is satisfied:
$$Q_{\alpha}(\pi)\geq \frac{\beta}{\beta - 1}k.$$



This entry rule provides a specific cutoff value determining the firm's decision to enter the market. This cutoff value depends on the shape $\beta$ and $\alpha$-quantile of the profit shock distribution along with the fixed entry cost.

The parameter $\beta$ is of particular interest, capturing a notion of skewness in the distribution. For high values of $\beta$, the distribution is characterized by long skinny tails. This parameter has a significant influence on the firm's decision-making process, as an optimistic firm will be more inclined to enter the market when there is even a slim chance of extremely high profits.

\subsection{Optimism and Risk Preferences}

In the previous section, we discussed the role of optimism on market entry. However, in the study of entrepreneurial endeavors, both risk preferences and optimism have received extensive attention. While optimism leads DMs to overweight the positive aspects of risk by focusing on high payoff outcomes, risk aversion reflects a tendency to prefer riskless prospects to risky ones. In this section, we demonstrate that our WT model can be extended to capture the separate effects of optimism and risk aversion, reflecting the complex interrelationship between these two. 

Many studies have tested whether individual differences in risk aversion explain entrepreneurial endeavors, and \cite{Parker_2009} provides a systematic review of these studies. The findings are inconclusive, with some finding risk tolerance to drive entrepreneurship (\cite{ahn2010attitudes}) while others do not (\cite{holm2013entrepreneurs}). Optimistic beliefs provide another plausible explanation for entrepreneurship (\cite{puri2007optimism}; \cite{bengtsson2014bright}). \cite{aastebro2014seeking} note that ``much of the research on entry into entrepreneurship has tended to focus on single factors...  (but) the time is ripe to compare and contrast these factors.'' With this in mind, we present a decision criterion that reflects both WT and risk aversion.

\smallskip

Similar to the utility function in Equation (\ref{Optimistic_RUM}), the DM's utility for actions $a\in A$ depends on a deterministic value $u(a)$ and an additively separable term dependent of random variable $\omega_a$. Let the random variable $\omega_a$ be defined by  $\omega_a \sim N(0,\omega_a^2)$. To capture risk aversion, we consider our DM to have an exponential utility function 

\begin{equation}\label{exponential}
u(a,\omega_a) = u(a) - e^{-r\omega_a},\end{equation}
which exhibits constant absolute risk aversion with corresponding risk-aversion parameter $r$. To focus on the case where the DM is risk averse, we restrict consideration to $r > 0$. We can formalize a decision rule for a DM who is both risk-averse and optimistic.

\begin{corollary}\label{wtriskaverse}
Consider a WT problem with the payoff structure as defined by Equation \eqref{exponential} with optimism level $\alpha$ and risk aversion parameter $r$. Then the WT agent solves the following problem:
\begin{equation*}
\max_{a\in A}\left\{u(a) - \frac{1}{(1-\alpha)}exp(\frac{1}{2}r^2 \sigma^2)\left(1- \Phi\left({r\sigma} + \frac{1}{\sigma}{Q_\alpha(\omega_a)}\right)\right)\right\}.
\end{equation*}
where $\Phi$ denotes the CDF function for the standard normal function

\end{corollary}

The decision-making criterion laid out in Corollary \ref{wtriskaverse} separately captures the effects of optimistic beliefs and risk aversion, highlighting each factor's role in shaping choice. A few points are in order on this criterion. Firstly, we see that the $\alpha$-quantile again plays a critical role in decision-making. As we have previously demonstrated, it captures a DM's degree of optimism, and this fact holds true under risk aversion.

Second, by applying Corollary \ref{WT_CvAR_Corollary}, we see that for $\alpha \rightarrow 0$, the criterion tends toward $\max_{a\in A}\{u(a) - exp(\frac{1}{2}r^2\sigma^2)\}$. Similarly, for $\alpha \rightarrow 1$, the criterion tends toward maximization of the deterministic portion: $\max_{a\in A}\{u(a)\}$. Taken together, these facts highlight that as a dm  becomes more optimistic, the influence of risk aversion diminishes, reducing the tendency to favor less risky actions. In the context of capital budgeting, \cite{gervais2002positive} similarly notes optimism's ability to offset risk aversion.

\smallskip

Finally, note that the formulation with normal error terms and CRRA utility allows us to yield a closed-form expression for the decision-making criterion. However, similar analysis is possible using numerical integration for other notions of risk aversion and random variables drawn from different distributions.

\section{Conclusions}\label{concusions} 
In this paper, we have developed a model of WT and optimism. Our key finding is that WT behavior can be expressed as equivalent to superquantile utility maximization. We achieve this by introducing the threshold beliefs distortion cost. From a behavioral perspective, this implies that an optimistic agent relies on the upper tail of the utility distribution and assesses the expected utility conditional on being above a specific quantile. Moreover, our analysis reveals that a WT agent's distorted beliefs overweight the probability associated with desirable outcomes by underweighting the probability of less desirable ones. Unlike wishful thinking based on the Kullback-Leibler divergence, this reweighting is done in a binary fashion.

Our model of optimistic decision-making based on censored beliefs yields numerous insights into the effects of WT. We formalize a connection between WT behavior and a preference for skewness, demonstrating that optimists focus on the shape of the right tail in their analysis. Furthermore, we demonstrate that optimistic beliefs can be easily implemented into random utility models. In particular, we provide a condition that ensures that stochastic choice is monotone, allowing one to test for changes in decision-making optimism levels. We then discuss how optimism can drive market entry and present a decision criterion that captures the complex relationship between optimistic beliefs and risk aversion.

Looking ahead, there are several avenues for extending and enriching our model, many involving the incorporation of WT into environments of information acquisition and stochastic choice.\footnote{\cite{strzalecki2025stochastic} presents a comprehensive study of information acquisition models and stochastic choice.} One direction involves investigating how WT behavior unfolds in a dynamic environment, potentially yielding new insights into the development of sentiments. Another involves embedding the censored beliefs model into a persuasion problem, as studied by \cite{augias2020persuading} with KL-based wishful thinking. Finally, a revealed-preferences approach may offer valuable perspectives on the choice data arising from censored beliefs, and the data-driven framework of \cite{caplin2015testable} could provide a promising direction for this work.


\bibliographystyle{plainnat} 
\bibliography{bibliography.bib}
\newpage
\section*{Appendix: Proofs}
\subsection*{Proof of Proposition \ref{dualprob}}

We offer two proofs of this result. First, we prove Proposition \ref{dualprob} using an argument that exploits the properties of the threshold beliefs distortion cost function without invoking convex duality.\footnote{We are very grateful to the AE for suggesting this simpler proof.} A subjective belief has cost zero if it is not too far from the prior; otherwise, it is infinity. To formalize this, notice that the cost can be slightly rewritten as:

 \begin{equation*}\label{Cost_phi}
   C_\alpha(G\| F)=\begin{cases}
	0, & 0\leq  g(\omega)\leq \frac{f(\omega)}{1-\alpha},  \; \forall \; \omega \in \Omega\\
	+\infty & \text{otherwise}
\end{cases}
\end{equation*}

As such, the subjective density can be at most $\frac{1}{1-\alpha}$ times larger than the prior. Then the interior maximization problem for action $a$ multiplying the density function $f$ of $(F)$ by $\frac{1}{1-\alpha}$ and selecting cutoff $\omega^*$ such that
$$\int_{\left\{\omega: u(a, \omega) \geq u\left(a, \omega^*\right)\right\}} \frac{1}{1-\alpha} d f=1
$$In this way, the DM maximizes her SEU by choosing an optimistic subjective belief, which assigns probability zero to states with utility lower than $\omega^*$, without suffering any $\operatorname{cost}\left(C_\phi=0\right.$.) Formalized, this yields expression (\ref{CVaR}).
\smallskip

Our second proof relies on convex duality arguments used in \cite{BurghMelo2023}. In particular, from Lemma 1 in \cite{BurghMelo2023}, the function $V_\alpha$ can be rewritten as:
$$V_\alpha(a) =\min_{\lambda_a\in[\underline{u}_a,\bar{u}_a]} \left\{ \lambda_a+\EE_{F}(\phi^\ast(u(a,\omega)-\lambda_a))\right\}$$
where $\phi^*(s) = \sup _{t \in \operatorname{int} \operatorname{dom} \phi}{s t-\phi(t)}$ is the convex conjugate of function $\phi$. It is straightforward to see that the convex conjugate of the function $\phi$ defined by Equation (\ref{Indicator_phi}) is given by $\phi^\ast(s)={1\over 1-\alpha}\max\{s,0\}$. Defining $s_a=u(a,\omega)-\lambda_a$ and  plugging in the convex conjugate $\phi^*$  we get: $$V_\alpha(a)=\min_{\lambda_a\in[\underline{u}_a,\bar{u}_a]}\left\{\lambda_a+{1\over 1-\alpha}\EE_F(\max\{u(a,\omega)-\lambda_a,0\})\right\}$$\eproof

\subsection*{Proof of Proposition \ref{WT_CvAR}} Let $F(a,\lambda_a)\triangleq \lambda_a+{1\over 1- \alpha}\EE_F(\max\{u(a,\omega)-\lambda_a,0\})$. Then  expression (\ref{CVaR}) can be rewritten as  

\begin{equation*}\label{CVAR_2}
V_{\alpha}(a)=\min_{\lambda_{a}\in \Lambda(a)}F(a,\lambda_a).
\end{equation*}

By \cite[Thm. 1]{Rockafellar2000OptimizationOC}, we know that $F(a,\lambda_a)$ is convex and continuously  differentiable with respect to $\lambda_a$. Thus  the optimal $\lambda_a^{\star}$ satisfies  the  necessary and sufficient first-order condition:
$$1-{ 1\over 1-\alpha}\mathbb{P}(u(a,\omega)\geq\lambda^\star_a)=0.$$
Rewriting the  previous expression, we get
$\alpha=\PP(u(a,\omega)\leq \lambda^\star_a)$. Then from Definition \ref{VaR_Definition} it follows that $\lambda_a^\star=Q _{\alpha}(a)$. This proves part (i). Part (ii) follows from using (i) combined with expression (\ref{CVaR}). To show (iii), we combine  Definition \ref{CvAR_Definition}  with the fact that $1-\alpha=\PP(u(a,\omega)\geq Q_{\alpha}(a))$. In particular, we can write the following:
$$
\textcolor{black}{
\begin{aligned} \bar{Q}_\alpha(a) & =Q_\alpha(a)+\frac{1}{1-\alpha} \int_{\Omega} \max \{u(a,\omega) -Q_\alpha(a),0\} f(\omega) d\omega \\ & =Q_\alpha(a)+\frac{1}{1-\alpha} \int_{\{\omega: u(a,\omega)\geq Q_\alpha(a) \}} u(a,\omega) f(\omega) d \omega-\frac{Q_\alpha(a)}{1-\alpha} \int_{\{\omega: u(a,\omega)\geq Q_\alpha(a) \}} f(\omega) d \omega \\ & =\frac{1}{1-\alpha}\int_{\{\omega: u(a,\omega)\geq Q_\alpha(a) \}}u(a,\omega) f(\omega) d \omega\\&=\EE_F(u(a,\omega)|u(a,\omega)\geq Q_\alpha(a)).
\end{aligned}}$$

Finally, part (iv) follows from the using the fact that $1-\alpha=\PP(u(a,\omega)\geq Q_\alpha(a)) $. Then, by using the definition of conditional distribution, the expression for  $G_{\alpha}(a)$ follows at once.\eproof
\subsection*{Proof of Corollary \ref{WT_CvAR_Corollary}}\textcolor{black}{ Part (i) is a direct application of Proposition 13 in \cite{ROCKAFELLAR20021443}. To show (ii), we note that under the assumption that $F_a$ is continuous and strictly increasing,  we know that $Q_\alpha(a)$ is uniquely defined. Then, by using the envelope theorem, we differentiate with respect to $\alpha$ to obtain the desired result. Part (iii) follows immediately from the (ii) which implies $\frac{\partial V_\alpha(a)}{\partial \alpha} \geq 0$. Part (iv) and (v) follows from the definition of the superquantile.} \eproof

\subsection{Proof of Proposition \ref{Prop_skewed_DC}}

The density function of the Pareto distribution is given by:

$$F(\omega_a) = 1 - (\frac{\bar{\omega}_a}{\omega_a})^{\beta_a}.$$

Because $F$ is a continuous distribution, the $\alpha$-quantile $Q_\alpha(a)$ must satisfy

$$\alpha = 1 - (\frac{\bar{\omega}_a}{Q_\alpha(a)})^{\beta_a}$$

Algebra along with Proposition \ref{WT_CvAR} (i) yields:

$$Q_\alpha(\omega_a) = \lambda_a^* = \bar{\omega}_a(1-\alpha)^{-1/\beta_a}$$

Now, we know that  $\bar{Q}_{\alpha}(\omega_a)={(1-\alpha)^{-1}}\int_{\alpha}^1 Q_\theta(\omega_a)d\theta$, and direct computation yields:
$$\bar{Q}_{\alpha}(\omega_a)=(1 - \frac{1}{\beta_a})Q_\alpha(\omega_a)$$

In this case, problem (\ref{wishful_thinking_problem}) can be expressed as:

$$\max_{a\in A}\left\{u(a)+ (1 - \frac{1}{\beta_a})Q_\alpha(\omega_a)\right\}$$

\eproof
\subsection{Proof of Proposition \ref{Monotone_RUM_Opt}} $\rho^{RUM}$ increasing $\Longrightarrow$ Eq. (\ref{Mon_Inequality}).  Assume $\rho^{RUM}$ is increasing in $\alpha$ for pair $(a,a^\prime)$. By definition, the probability of choosing $a$ over $a^\prime$ corresponds to $\rho^{RUM}_\alpha(a,a^\prime) = \PP(V_\alpha(a)+\epsilon_a\geq V_\alpha(a^\prime)+\epsilon_{a^\prime}) = \PP(V_\alpha(a)-V_\alpha(a^\prime) \geq \epsilon_a-\epsilon_{a^\prime})$.

We note that distribution of the  difference  $\epsilon_a-\epsilon_{a^\prime}$ is a continuous random variable with corresponding CDF $\Psi$ and density function $\psi$.  In particular, we can write$\rho^{RUM}_\alpha(a,a^\prime) = \Psi(V_\alpha(a)-V_\alpha(a^\prime))$. Then, using Corollary \ref{WT_CvAR_Corollary}(ii) yields: 
{\footnotesize $$ \frac{\partial \rho^{RUM}_\alpha(a,a^\prime)}{\partial \alpha} ={\psi(\Delta V_\alpha(a,a^\prime))\over (1-\alpha)^2}(\EE_{F}(\max\{ u(a,\omega)-Q_\alpha(a),0\})- \EE_{F}(\max\{u(a^\prime,\omega)-Q_\alpha(a^\prime),0\})\geq 0,$$}
where $\Delta V_\alpha(a,a^\prime)\triangleq V_\alpha(a)-V_\alpha(a^\prime)$.
Given that  $\psi$ is a positive density, we know t${\psi(V_\alpha(a)- V_\alpha(a^\prime))\over (1-\alpha)^2} > 0$  must hold:
$$\EE_{F}(\max\{ u(a,\omega)-Q_\alpha(a),0\})\geq \EE_{F}(\max\{u(a^\prime,\omega)-Q_\alpha(a^\prime),0\})\quad\forall \alpha\in(0,1).$$\\
Eq. (\ref{Mon_Inequality}) $\Longrightarrow \rho^{RUM}$ is monotone. Assume $\EE_{F}(\max\{ u(a,\omega)-Q_\alpha(a),0\})\geq \EE_{F}(\max\{u(a^\prime,\omega)-Q_\alpha(a^\prime),0\}) \geq 0$ holds for all  $\alpha \in (0,1)$.

We  now show that The CDF $\Psi$ characterizing the choice of $a$ over $a^\prime$  must be increasing when Eq. (\ref{Mon_Inequality}) holds. In particular, we know $\rho^{RUM}(a,a^\prime)=\Psi(V_\alpha(a)-V_\alpha(a^\prime))$. Then computing ${\partial \rho^{RUM}(a,a^\prime)\over \partial \alpha}$  we find that:

{\footnotesize $${\partial \rho^{RUM}(a,a^\prime)\over \partial \alpha}={\psi(\Delta V_\alpha(a,a^\prime))\over (1-\alpha)^2}(\EE_{F}(\max\{ u(a,\omega)-Q_\alpha(a),0\})- \EE_{F}(\max\{u(a^\prime,\omega)-Q_\alpha(a^\prime),0\})\geq 0,$$}
with $\Delta V_\alpha(a,a^\prime)=V_\alpha(a)-V_{\alpha}(a^\prime).$ Given that the term $\psi(\Delta V_\alpha(a,a^\prime))\over (1-\alpha)^2>0$,  $\rho^{RUM}(a,a^\prime)$ being monotone implies that  for all $\alpha\in (0,1)$ the following inequality must hold:
$$\EE_{F}(\max\{ u(a,\omega)-Q_\alpha(a),0\})\geq \EE_{F}(\max\{u(a^\prime,\omega)-Q_\alpha(a^\prime),0\}) \geq 0 ).$$
\eproof

\subsection{Proof of Corollary \ref{wtriskaverse}}

We have defined the random variable influencing payoffs by $\omega_a \sim N(0,\omega_a^2)$, and following our previous notation $\omega_a$'s $\alpha$-quantile as $Q_\alpha(\omega)$. From here, we can define the random variable of potential utility outcomes corresponding with $u(a,\omega_a) = u(a) - e^{-r\omega_a}$. We denote the density function of this distribution as $P_a$, and note that it is continuous and strictly increasing in $\omega_a$. Note that the WT value associated with action $a$ is given by:

\begin{multline*}
V_\alpha(a) = \EE_{P_a}(u(a,\omega)|\omega \geq Q_\alpha(\omega))\\ = \frac{1}{1-\alpha}\int_{Q_\alpha(\omega)}^{\infty} \frac{1}{\sigma \sqrt{2\pi}}exp(-\frac{x^2}{2\sigma^2})(u(a) - exp(-rx))dx\\
= q(a) - \frac{1}{1-\alpha}\int_{Q_\alpha(\omega)}^{\infty} \frac{1}{\sigma \sqrt{2\pi}}exp(-\frac{x^2}{2\sigma^2} - rx)dx\\
=q(a) - \frac{1}{1-\alpha}\frac{1}{2} \left[ \operatorname{erf}\left(\frac{r \sigma^2+ x}{\sqrt{2} \sigma}\right) \exp \left(\frac{1}{2} r^2 \sigma^2\right) \right]^\infty_{Q_\alpha(\omega)}\\
=q(a) - \frac{1}{1-\alpha}\frac{1}{2}exp \left(\frac{1}{2} r^2 \sigma^2\right)(1 - \operatorname{erf}\left(\frac{ r \sigma^2+ x}{\sqrt{2} \sigma}\right))
\end{multline*}

Finally, making use of the fact that $\operatorname{erf}(z/\sqrt{2}) = 2 \Phi(z) - 1$, we get:

$$V_\alpha(a) = u(a) - \frac{1}{(1-\alpha)}exp(\frac{1}{2}r^2 \sigma^2)\left(1- \Phi\left({r\sigma} + \frac{1}{\sigma}{Q_\alpha(\omega_a)}\right)\right)$$

So the WT agent with optimism parameter $\alpha$ solves:

\begin{equation*}
\max_{a\in A}\left\{u(a) - \frac{1}{(1-\alpha)}exp(\frac{1}{2}r^2 \sigma^2)\left(1- \Phi\left({r\sigma} + \frac{1}{\sigma}{Q_\alpha(\omega_a)}\right)\right)\right\}.
\end{equation*}

\section*{Online Material: Closed-form expressions for $\EE_{F_a}(\omega_a|\omega_a\geq Q_{\alpha}(\omega_a))$}\label{Closed_Form_CVaR}
In this section we describe several distributions that yield closed-form expressions for $\EE_{F_a}(\omega_a|\omega_a\geq Q_{\alpha}(\omega_a))$. In doing so, we use the results in \cite{Uryasev_et_al_2018}.

\subsection{The Logistic distribution}  For each $a\in A$, assume that  $\omega_a \sim \operatorname{Logistic}(\mu_a, s_a)$. Setting $\mu_a=0$ and $s_a>0$, for all $a\in A$, we obtain  $\EE_{F_a}(\omega_a)=0$ and $\mathbb{V}(\omega_a)=\frac{ s_a\pi^2}{3}$.  Accordingly, 
$$F_a(\omega_a)=\frac{1}{1+e^{-{\omega_a\over s_a}}},$$

Then $Q_\alpha(\omega_a)= \ln \left(\frac{\alpha}{1-\alpha}\right).  $ From Proposition \cite[Prop. 10]{Uryasev_et_al_2018} we know that 
$$\EE_{F_a}(\omega_a|\omega_a\geq Q_{\alpha}(\omega_a))=s_a\frac{ H(\alpha)}{1-\alpha}$$
where $H(\alpha)\triangleq-\alpha \ln (\alpha)-(1-\alpha) \ln (1-\alpha).$ Thus the problem (\ref{WT_RUM_sol}) can be expressed as:
$$\max_{a\in A}\left\{u(a)+s_a\frac{ H(\alpha)}{1-\alpha}\right\}.$$

\subsection{The Student-t distribution} Assume $\omega_a \sim$ Student $-t(\nu_a, s_a, \mu_a)$. where $\nu_a>0, s_a>0$, $\mu_a>0$ with $\EE_{F_a}(\omega_a)=\mu_a$ and $\mathbb{V}(\omega_a)=\frac{s_a^2 \nu_a}{v_a-2}$. Setting $\mu_a=0$, the Student $t$ distribution corresponds to

$$F_a(\omega_a)=1-\frac{1}{2} \mathcal{I}_{g(\omega_a)}\left(\frac{\nu_a}{2}, \frac{1}{2}\right)$$
where $f_a(\omega_a)=\frac{\nu_a}{\frac{\omega_a}{s}+\nu_a}, \mathcal{I}_t(a, b)$ is the regularized incomplete Beta function, and $\Gamma(a)$ is the Gamma function. 
\smallskip

From \cite[Prop. 12]{Uryasev_et_al_2018}, we know that 
$$\EE_{F_a}(\omega_a|\omega_a\geq Q_\alpha(\omega_a))=s_a\left(\frac{\nu_a+T^{-1}(\alpha)^2}{(v-1)(1-\alpha)}\right) l\left(T^{-1}(\alpha)\right)$$
where $T^{-1}(\alpha)$ is the inverse of the standardized Student-t cumulative distribution function and $l(\cdot)$ is the standardized Student-t probability density function.
\smallskip

Accordingly, the DM's problem (\ref{WT_RUM_sol}) can be rewritten as:
$$\max_{a\in A}\left\{u(a)+s_a\left(\frac{\nu_a+T^{-1}(\alpha)^2}{(v-1)(1-\alpha)}\right) l\left(T^{-1}(\alpha)\right)\right\}.$$
\subsection{The generalized extreme value distribution} Next, we discuss the generalized extreme value (GEV) distribution case. Formally, we  assume that $\omega_a$ follows a GEV distribution, which we denote as $\omega_a \sim GEV(\mu_a, s_a, \xi_a)$. Recall that GEV parameters have range $\mu_a \in \mathbb{R}, s_a>0, \xi_a \in \mathbb{R}$. The parameters $\mu_a,s_a,\xi_a$ capture
location, scale, and shape respectively. 
The cumulative distribution corresponds to:
$$F_a(\omega_a)=\left\{\begin{array}{ll}
e^{-\left(1+\frac{\xi_a(\omega_a-\mu_a)}{s_a}\right)^{\frac{-1}{\xi_a}}} & \xi_a \neq 0, \\
e^{-e^{-\left(\frac{\omega_a-\mu_a}{s_a}\right)}} & \xi_a =0
\end{array},\right.$$
From \cite[Prop. 15]{Uryasev_et_al_2018} we know that 

\begin{equation*}\label{Eq_GEV}
\EE_{F_a}(\omega_a|\omega_a\geq Q_\alpha(\omega_a))= \begin{cases}\mu_a+\frac{s_a}{\xi_a(1-\alpha)}\left[\Gamma_L\left(1-\xi_a, \ln \left(\frac{1}{\alpha}\right)\right)-(1-\alpha)\right] & \xi_a \neq 0 \\ \mu_a+\frac{s_a}{(1-\alpha)}(y+\alpha \ln (-\ln (\alpha))-\operatorname{li}(\alpha)) & \xi_a=0\end{cases}
\end{equation*}
where $\Gamma_L(a, b)=\int_0^b p^{a-1} e^{-p} d p$ is the lower incomplete gamma function, $\operatorname{li}(x)=$ $\int_0^\alpha \frac{1}{\ln p} d p$ is the logarithmic integral function, and $y$ is the Euler-Mascheroni constant.
\smallskip

Using the expression above we can rewrite the DM's problem (\ref{WT_RUM_sol}) accordingly.

\subsection{The Generalized Pareto Distribution}

Next, we discuss the generalized extreme value (GEV) distribution case. Formally, we  assume that $\omega_a$ follows a GEV distribution, which we denote as $\omega_a \sim GPD(\xi_a, \beta_a)$. The GPD is characterized by two parameters: $\xi_a \in \mathbb{R}$ and $\beta_a > 0$, and the density function can be expressed as follows:
\begin{equation*}\label{Generalized_Pareto}
F_{a}(\omega_a)= \begin{cases}1-\left(1+\frac{\xi_a \omega_a}{\beta_a}\right)^{-1 / \xi_a} & \xi_a \neq 0, \\ 1-\exp \left(-\frac{\omega_a}{\beta_a}\right) & \xi_a=0\end{cases}
\end{equation*}
where $\omega_a\in[0, \infty)$ for $\xi_a \geq 0$ and $\omega_a \in[0,-\beta_a / \xi_a]$ for $\xi_a<0$.

The parameter $\beta_a$ acts to determine the scale of the distribution, while $\xi_a$ determines the shape. In particular, when $\xi_a = 0$, the GPD reduces to an exponential distribution. When $\xi_a > 0$, $F_{a}$ represents a Pareto distribution as we explore in section \ref{skewedsection}, albeit with a slightly different parametrization. For the GPD with $\xi_a > 0$, the $k$th moment does not exist when $k \geq 1 / \xi_a$, similar to the case of the Fréchet distribution. Critically, for $\xi_a \geq 1$, the first moment does not exist and as such the superquantile is not defined. On the other hand, when $\xi_a < 0$, the density $F_a$ yields the Pareto Type II distribution. For $\xi_a >1$, $\EE(\omega_a) = \infty$, so we  focus on the case where $\xi_a \in (-\infty,0) \cup (0,1)$. The optimal multiplier is given by
\smallskip

$$\lambda^*_{a}=\frac{\beta_a}{\xi_a}((1-\alpha)^{-\xi_a} -1).$$

By one definition of the superquantile we know that  $\bar{Q}_{\alpha}(a)={(1-\alpha)^{-1}}\int_{\alpha}^1 Q_\theta(a)d\theta$, direct computation yields:
$$\bar{Q}_{\alpha}(a)={Q_{\alpha}(\omega_a)\over 1-\xi_a}+{\beta_a\over \xi_a}((1-\xi_a)^{-1} - (1-\alpha)^{-1}).$$

Using the expression above we can rewrite the DM's problem (\ref{WT_RUM_sol}) accordingly. And similar analysis yields the DM's problem when $\xi_a = 0$.

\end{document}